\documentclass{hotnets22}

\usepackage{times}  
\usepackage{hyperref}
\usepackage{titlesec}

\hypersetup{pdfstartview=FitH,pdfpagelayout=SinglePage}

\usepackage[linesnumbered,ruled,vlined]{algorithm2e}
\usepackage{listings}

\usepackage{tikz}
\usepackage{amsmath}
\usepackage{filecontents}

\usepackage{comment}
\usepackage[shortlabels]{enumitem}
\newenvironment{myitemize}
{ \begin{itemize}[leftmargin=0.2in]	
		\vspace{-1ex}	
		\setlength{\itemsep}{2pt}
		\setlength{\parskip}{1pt}
		\setlength{\parsep}{1pt}    }
	{ 	 \end{itemize}                    }

\newenvironment{myenumerate}
{ \begin{enumerate}[leftmargin=0.2in]
		\vspace{-1ex}
		\setlength{\itemsep}{2pt}
		\setlength{\parskip}{1pt}
		\setlength{\parsep}{1pt}    }
	{ \end{enumerate}                  }
	
\usepackage[linesnumbered,ruled,vlined]{algorithm2e}
\usepackage{tabularx}
\usepackage{booktabs}

\usepackage{algorithmic}
\usepackage{graphicx}
\usepackage{textcomp}
\usepackage{xcolor}
\usepackage{amsfonts}

\usepackage{graphicx}
\usepackage{float}
\usepackage{subfig}

\usepackage{url}
\usepackage{multirow}

\usepackage{hyphenat}
\usepackage{comment}

\newcommand{\system}{\texttt{TransNFV}\xspace}

\newcommand{\zhonghao}[1]{\textcolor{blue}{#1}}

\usepackage{tikz}
\newcommand*\circled[1]{\tikz[baseline=(char.base)]{
            \node[shape=circle,fill,inner sep=1pt] (char) {\textcolor{white}{#1}};}}
\usepackage{makecell}

\newcommand{\compact}{\vspace{-0pt}}

\usepackage{float}
\begin{document}



\title{TransNFV: Integrating Transactional Semantics for Efficient State Management in Virtual Network Functions}

\author{Zhonghao Yang \\ SUTD \and Shuhao Zhang \\ SUTD \and Binbin Chen \\ SUTD
}

\maketitle

\begin{abstract}
Managing shared mutable states in high concurrency state access operations is a persistent challenge in Network Functions Virtualization (NFV). This is particularly true when striving to meet chain output equivalence (COE) requirements. This paper presents \system, an innovative NFV framework that incorporates transactional semantics to optimize NFV state management. The \system integrates VNF state access operations as transactions, resolves transaction dependencies, schedules transactions dynamically, and executes transactions efficiently. Initial findings suggest that \system maintains shared VNF state consistency, meets COE requirements, and skillfully handles complex cross-flow states in dynamic network conditions. \system thus provides a promising solution to enhance state management and overall performance in future NFV platforms.

\end{abstract}

\compact
\section{Introduction}

Network Function Virtualization (NFV) has revolutionized network function deployment through software-based Virtualized Network Functions (VNFs)~\cite{mijumbi2015network}. This shift brings improved flexibility, adaptability, and cost-efficiency compared to traditional hardware-based solutions. However, managing stateful VNFs, particularly those handling cross-flow states shared among multiple instances, poses considerable challenges. Moreover, NFV must fulfil chain output equivalence (COE) requirements, ensuring that the VNF actions across an NFV chain align with a hypothetical, always-available, single NF with infinite capacity~\cite{khalid2019correctness}. These challenges are even more pronounced in emerging 5G/6G networks, where the dynamic bandwidth allocation model based on user needs and network conditions necessitates optimal shared resource allocation and efficient VNF performance~\cite{barakabitze20205g}.

Despite significant progress, existing NFV frameworks face difficulties in managing shared VNF states and complying with COE requirements under dynamic network conditions~\cite{bremler2016openbox, meng2019micronf, palkar2015e2, rajagopalan2013split, gember2014opennf, khalid2019correctness, woo2018elastic}. To address this, we present \system, a ground-breaking NFV framework designed to manage cross-flow states efficiently while ensuring COE compliance. \system distinguishes itself by employing database transaction concepts to encapsulate state access operations in VNFs into distinct units. This unique approach ensures the consistency and reliability of shared mutable states across VNF instances and provides an effective strategy for managing cross-flow states while addressing COE complexities.

To realize such an approach, \system introduces four novel mechanisms: (i) transaction modelling for VNFs, (ii) transaction planning to resolve dependencies, (iii) adaptive transaction scheduling based on real-time execution status, and (iv) efficient concurrent transaction execution. These features contribute to \system's scalability, adaptability, and robustness. Initial experiments demonstrate the potential of \system to significantly improve strategies in managing cross-flow states under dynamic network conditions. In particular, our results show that \system outperforms the most closely related work, CHC framework~\cite{khalid2019correctness}, with up to double the throughput and 2.5 times lower processing latency while maintaining the same COE compliance.

The remainder of this paper is organized as follows: Section~\ref{sec:motivation} presents a motivating example to underscore the need for an innovative approach to shared state management in NFV. Section~\ref{key_mechanisms} details the unique mechanisms of \system. Section~\ref{prototype} provides an in-depth view of the \system prototype. Section~\ref{Experiments} discusses our early-stage results, and we conclude in Section~\ref{conclusion} with a summary and discussion of potential future work.
\compact
\section{Motivation}
\label{sec:motivation}
Achieving \textit{chain output equivalence} (COE) in NFV chains, especially under high-throughput, low-latency demand is a daunting task. Existing NFV frameworks fall short of fully addressing these challenges. This section underscores these hurdles via an example.

\subsection{Motivating Example}
\label{subsec:example}
\begin{figure}
    \includegraphics[width=\linewidth]{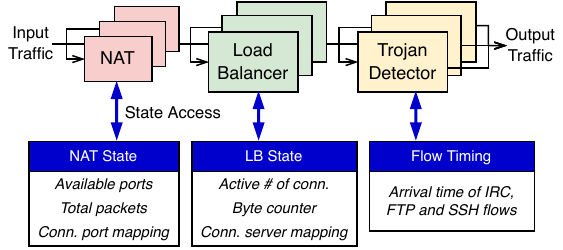}
    \caption{An Example NFV Chain.}
    \label{fig:SFC}
\end{figure}
Figure~\ref{fig:SFC} shows a representative NFV chain that consists of Network Address Translation (NAT), Load Balancing (LB), and Trojan Detection (TD). Although these functions, deployed sequentially across different Virtual Network Function (VNF) instances, handle distinct packet flows, they rely on shared state objects, which adds complexity to state management in VNFs~\cite{de2014beyond}. Each function operates as follows:

\begin{myenumerate}
\item \textbf{Network Address Translation (NAT):} On receiving a new connection, NAT instances select an available port from a shared list and update the per-flow state with the connection's port mapping. Simultaneously, they manage cross-flow states like the available ports and total packet count.
\item \textbf{Load Balancing (LB):} LB instances allocate incoming packets to a suitable server based on its current capacity. They also maintain a connection-to-server mapping and collaborate with other instances to keep track of the active connections and byte count per server.
\item \textbf{Trojan Detection (TD):} TD instances monitor flow timings across different connections on each host to detect Trojan Horse attacks by identifying predefined malicious patterns.
\end{myenumerate}

From this operational breakdown of the representative NFV chain, we distill two overarching challenges intrinsic to the management of states in VNFs. The first challenge arising from this example is the need for COE in NFV chains, a concept elaborated in~\cite{khalid2019correctness}. This requirement equates to:
\begin{myitemize}
\item \textbf{Availability}:
Every instance in our chain, whether NAT, LB, or TD, must have continuous and up-to-date access to the shared state. This need is crucial even in the face of network fluctuations, such as instance failures or changes in traffic allocation. For instance, if a NAT instance fails, the system should swiftly provide an up-to-date state to the replacement instance to ensure seamless service.
\item \textbf{Consistency}:
Certain instances in our example, like NAT and LB, need to update shared states, including available ports and the active number of connections per server. Synchronizing these concurrent state updates is crucial for maintaining the consistency of shared cross-flow states.
\item \textbf{Ordering}:
In some network functions, it's vital to track the order of incoming traffic. For instance, a TD instance looks for threats based on a specific sequence of network events. Any mix-up in this order, possibly due to network congestion or recovery from failure, can cause incorrect detections. The issue gets more complex with multiple detectors, underscoring the importance of preserving traffic order in managing VNF states.
\item \textbf{Isolation}:
Every module in the NFV chain should be able to recover independently from a failure without causing disruptions in the chain. For instance, if an LB instance fails and recovers, it should not affect the operation of NAT or TD instances.
\end{myitemize}

The second challenge resides in the need to maintain execution efficiency while ensuring COE. Particularly in our example, this involves several key components:
\begin{myitemize}
    \item \textbf{Concurrent State Access}: Both NAT and LB instances require concurrent reading and updating of shared state objects across multiple instances, such as the available ports list and server capabilities. The absence of adequate synchronization mechanisms could lead to significant contention overhead, affecting efficiency;
    \item \textbf{Dynamic Traffic Conditions}: The NFV Chain must operate under dynamic traffic conditions characterized by frequent instance failures and network bursts. This requires mechanisms capable of discarding associated updates, reverting to a safe state, and efficiently redistributing traffic processing to prevent redundancy. For instance, a sudden surge in traffic could overwhelm the LB instances, necessitating rapid reallocation of traffic to maintain service quality;
    \item \textbf{Cross-flow State Management}: The complexity of managing cross-flow states under such dynamic conditions is a non-trivial task. Given the shared state nature of our example, i.e., available ports, server capabilities, and active connections per server maintaining efficiency while ensuring COE is an intricate challenge.
\end{myitemize}
\subsection{Limitations of Previous Work}
\label{subsec:limitations}
The field of Network Function Virtualization (NFV) has witnessed considerable progress. Yet, current NFV frameworks~\cite{sherry2015rollback, rajagopalan2013pico, gember2014opennf, khalid2019correctness, woo2018elastic, rajagopalan2013split} struggle to efficiently manage VNF states while maintaining \textit{chain output equivalence} (COE).

Certain frameworks, such as FTMB~\cite{sherry2015rollback}, Pico Replication~\cite{rajagopalan2013pico}, and Split/Merge~\cite{rajagopalan2013split}, primarily focus on state access availability. They seek to preserve state availability during traffic reallocation and failure recovery by discarding state changes. However, their assumption of static states restricts their applicability in real-world situations, which commonly involve state changes.

Managing shared mutable VNF states and satisfying COE requirements is a daunting task. Frameworks like OpenNF~\cite{gember2014opennf} and S6~\cite{woo2018elastic} have attempted to address mutable states within a single VNF but not across multiple instances. Other frameworks~\cite{bremler2016openbox, meng2019micronf, palkar2015e2, rajagopalan2013split, gember2014opennf, khalid2019correctness, woo2018elastic} have made strides to support mutable states across multiple VNF instances, but they either overlook cross-flow states or make assumptions of perfect traffic partitioning, which simplifies the issue but neglects the intricacies of state sharing among instances.

Recently, certain frameworks, such as OpenNF~\cite{gember2014opennf}, CHC~\cite{khalid2019correctness}, and others~\cite{woo2018elastic, rajagopalan2013split}, have strived to maintain cross-flow state consistency during execution. However, they still face difficulties with the rigorous demands of COE, like state consistency during instances' replication and ensuring state access order. These solutions also tend to fall short in dynamic network conditions, including instance failure or traffic reallocation.

Among these, the CHC framework~\cite{khalid2019correctness} stands out for its attempt to fully accommodate COE requirements. Yet, it also has limitations. Its coarse-grained approach to managing cross-flow state results in considerable overhead. CHC's strategy for caching or flushing state objects, based on upstream traffic partitioner information, leads to frequent transfers between cache and main memory, resulting in high communication overhead. Moreover, the employment of mechanisms like locks to preserve consistency across multiple instances incurs additional overhead.

\compact
\section{Key Mechanisms}
\label{key_mechanisms}
\system combines transactional semantics with NFV state management, enabling efficient stateful VNFs execution while ensuring \textit{chain output equivalence} (COE). This is achieved through four key mechanisms: 1) modelling shared state access operations of VNFs as transactions, 2) identifying and resolving dependencies among VNF state access operations using a task precedence graph (TPG), 3) adaptively scheduling transactional workloads based on real-time execution status, and 4) execute transactions correctly and efficiently. These approaches help \system tackle challenges associated with shared state management in previous NFV frameworks.

\subsection{Modelling VNF State Access}
\label{subsec:transaction_modelling}
\system employs a graph-based representation to express the VNF chain as a directed acyclic graph (DAG), where vertices represent VNFs, and edges denote the flow of traffic between VNFs. Additionally, \system outlines three types of state access operations in each VNF: $READ$, which fetches the value of a state object; $WRITE$, which updates the state object with specific values; and $READ\_MODIFY$, which reads and modifies the value of a state object. \system models VNF execution logic through three steps: pre-processing, state access, and post-processing. 

\begin{myitemize}
\item The \emph{pre-processing} step includes identifying state entries linked to state access requests of incoming packets, as well as executing parts of the VNF logic that do not require accessing the state.
\item The \emph{state access} step involves state access operations (i.e., $READ$, $WRITE$ or $READ\_MODIFY$) on the shared state objects. 
\item The \emph{post-processing} step performs additional actions based on the results of the state access operations, and carries out parts of the VNF logic that rely on the returned value of state access.
\end{myitemize}

To ensure VNF state consistency and integrity, \system incorporates transactional semantics into VNF state management. Notably, it treats a set of state access operations collected during the \textit{state access} step as a single atomic transaction, meaning all modifications triggered by a packet are either fully committed or completely aborted.
This approach provides an efficient mechanism to handle VNF instance failures. When a VNF instance fails, \system ensures any modifications made by the failed instance are discarded and isolated from other operations. Furthermore, the failed state access operations can be re-executed as a transaction to ensure correct execution, leading to reliable and consistent state objects even under dynamic network conditions.

\subsection{Dependency Identification}
\label{subsec:transaction_dependency}
\system adapts to the dynamic scheduling of transactional workloads at runtime by mapping these workloads and their dependencies to a task precedence graph (TPG). The TPG is designed to capture and resolve three types of dependencies:

\begin{myitemize}
\item \emph{Temporal Dependency (TD)}: This depicts the temporal relationship between two state access operations stemming from different transactions, but targeting the same state. TD ensures that state accesses maintain an order that aligns with the sequence of packet arrivals, thereby satisfying the \textit{ordering} requirement.
\item \emph{Parametric Dependency (PD)}: This captures the parametric relationship between two state access operations that write to the same state, where one write operation is contingent on the outcome of the other. PD tracking is employed by the system to address potential conflicts among write operations in the VNF logic, adhering to the \textit{consistency} requirement.
\item \emph{Logical Dependency (LD)}: This denotes the logical relationship among state access operations within a single transaction. By tracking LD, \system ensures that all operations within a transaction are tightly connected and, in the event of failures, can be collectively aborted, adhering to the \textit{availability} and \textit{isolation} requirements.
\end{myitemize}

\system captures the transactional dependencies within each batch of state transactions by constructing a TPG. However, identifying these dependencies can pose challenges due to the potential \textit{out-of-order arrival} of network packets. To address this, \system divides the TPG construction process into two primary phases: the packet processing phase and the state access processing phase. These phases are alternated periodically, allowing for effective management of different packet batches. The period can be fine-tuned to balance execution latency and throughput.

\begin{myitemize}
\item \textit{Packet Processing Phase}: During this phase, LDs within the same transaction are identified based on their statement order. To identify TDs, operations are sorted by timestamp and inserted into key-partitioned concurrent lists corresponding to each operation's target state. For operations with PD, 'proxy' operations are inserted into the lists. These act as placeholders for the potential reading of a state object that may be modified by another operation. \item \textit{State Access Processing Phase}: During this phase, all subsequent state transactions are halted until \system reverts to the packet processing phase. Here, TDs and PDs are efficiently identified by iterating through the sorted lists and the 'proxy' operations, respectively. This efficient identification of dependencies and swift TPG construction are crucial for \system to adapt its scheduling strategy to the varying workload characteristics of different state transaction batches.
\end{myitemize}

\subsection{Transaction Scheduling}
\label{subsec:transaction_scheduling}
As part of its execution process, VNF instances use the constructed TPG to guide the dynamic scheduling of transactional workloads for concurrent processing. This involves exploring the TPG to identify executable tasks (i.e., state access operations) while respecting dependencies among tasks and attempting to maximize the concurrent execution. To adapt to varying workload characteristics and optimize opportunities for parallelism, \system employs different scheduling strategies. These strategies are determined in real-time based on three key dimensions:
\begin{myenumerate}
    \item \emph{Exploration of Remaining State Access Operations}: This dimension pertains to how the TPG is navigated during the scheduling process. VNF instances can employ either a structured approach, such as depth-first or breadth-first search-like traversal, or an unstructured approach that relies on random traversal. Depth-first and breadth-first approaches prioritize tasks that are deeper or wider in the graph, respectively. On the other hand, the random traversal approach does not favour any particular depth or breadth of tasks. The choice of the traversal method affects how soon certain tasks are scheduled and hence can impact the overall parallelism.
    \item \emph{Scheduling Granularity}: This dimension pertains to the size of the task unit that is scheduled at one time. During the exploration process, VNF instances can decide to schedule either a single operation or a group of operations as a unit of scheduling. Scheduling a single operation may reduce complexity but also limit parallelism. In contrast, scheduling a group of operations can increase parallelism but may also increase the complexity of handling dependencies among operations.
    \item \emph{Transaction Abort Handling}: This dimension pertains to how failed operations are handled. VNF instances can adopt an eager abort approach, where a failed operation is aborted immediately, thereby minimizing computation wastage for other operations that would have been dependent on the failed operation. However, this approach may incur high context-switching overhead. Alternatively, a lazy abort approach can be adopted, where failed operations are allowed to continue until a convenient point, reducing context-switching overhead at the risk of potentially wasting computational resources on tasks that will eventually be aborted.
\end{myenumerate}

The selection of a strategy for each dimension depends on the unique characteristics of the workload (such as the distribution of dependencies among operations) and the specific requirements of the application (such as tolerance for wasted computation or context-switching overhead). To ensure efficient and effective workload scheduling, \system may employ a decision model, which is currently under investigation, that dynamically determines the most suitable strategy for each dimension at runtime, based on current and projected workload characteristics. 

\subsection{Transaction Execution}
\label{subsec:transaction_execution}
The execution process within the \system framework seamlessly integrates the TPG, individual finite state machines for each state access operation, and a multi-versioning state table. This integration ensures robust handling of transactional workloads and guarantees system consistency and transactional accuracy.

Each vertex of TPG is annotated with a \textit{finite state machine} that represents its current status. The state machine transitions between various states reflecting the scheduling and execution processes: 1) \emph{Blocked}: The operation is pending execution due to unresolved dependencies; 2) \emph{Ready}: All dependencies have been resolved, and the operation is poised for execution; 3) \emph{Executed}: The operation has been processed successfully; 4) \emph{Aborted}: The operation has been terminated due to processing failures either from itself or its dependent operations.

To complement the TPG, \system uses a multi-versioning state table to manage consistency during concurrent state access operations. The table maintains a mapping between state access timestamps and state versions, associating every state access operation with a unique state version. This strategy safeguards system consistency, preventing the impact of operation failures or aborts from propagating throughout the system.

In addition, the multi-versioning state table provides a robust recovery mechanism in the face of network or instance failures. By facilitating the rollback of failed transactions to a safe stage, \system can recover states without compromising system consistency or integrity. This meticulous strategy enhances \system's capability to efficiently manage a diverse range of workloads.


\compact

\section{System Prototype}
\label{prototype}
\begin{figure}
\centering
\begin{minipage}{1.05\linewidth}
\includegraphics[width=\linewidth]{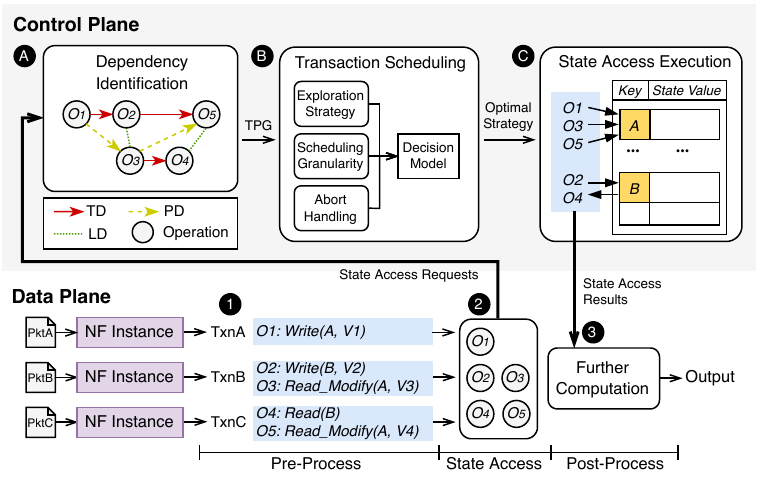}
    \caption{TransNFV System Prototype.}
    \label{fig:Prototype}    
\end{minipage}    
\end{figure}

\system's architecture divides control plane and data plane functionalities. The control plane manages the VNF state, while the data plane handles VNF execution tasks such as packet reception, forwarding, and computations based on state access results. We explain the \system workflow in Figure~\ref{fig:Prototype} in six steps.


\textbf{\system in the data plane.} 
The data plane incorporates packet reception, forwarding, and NF-specific computations. We enhance \textit{LibVNF}~\cite{LibVNF}, an open-source VNF library, to support data plane functions in \system. Although the libVNF API is versatile enough to build various VNFs and manage states across multiple VNF replicas, it lacks complete COE compliance, such as consistency and order requirements. We extend libVNF using the three-step transactional model outlined in Section~\ref{subsec:transaction_modelling}:
\textbf{\circled{1}} \textit{Pre-processing:} Upon packet reception, VNF instances formulate the corresponding state access operations into a transaction, maintaining system consistency.
\textbf{\circled{2}} \textit{State Access:} After constructing the transactions, they are sent to the control plane, where execution suspends until state access results are received, ensuring synchronized state access across VNF instances.
\textbf{\circled{3}} \textit{Post-processing:} Once state access results are acquired, NF-specific computations proceed, the output of which is used for further processing or response generation.

\textbf{\system in the control plane.} 
The control plane ensures efficient state access operations and consistency of NFV states. It is made up of three key components:
\textbf{\circled{A}} \textit{Dependency Identification:} This identifies transactional dependencies present within the incoming batch of state transactions, as discussed in Section~\ref{subsec:transaction_dependency}, allowing for efficient and consistent processing.
\textbf{\circled{B}} \textit{Transaction Scheduling:} Once the stateful Transaction Processing Graph (TPG) is formed, the scheduler analyses the workload characteristics, and guided by a heuristic decision model, determines optimal scheduling decisions, as detailed in Section~\ref{subsec:transaction_scheduling}.
\textbf{\circled{C}} \textit{Transaction Execution:} Executor threads implement state access operations as per the scheduling decisions, maintaining correctness, as detailed in Section~\ref{subsec:transaction_execution}. In case of potential aborts or failures, the multi-versioning state tables allow the recovery of states, preserving system integrity and consistency.

\begin{figure*}[h]
\centering
\begin{minipage}{\textwidth}
\subfloat[Throughput]{
\includegraphics[width=0.33\textwidth]{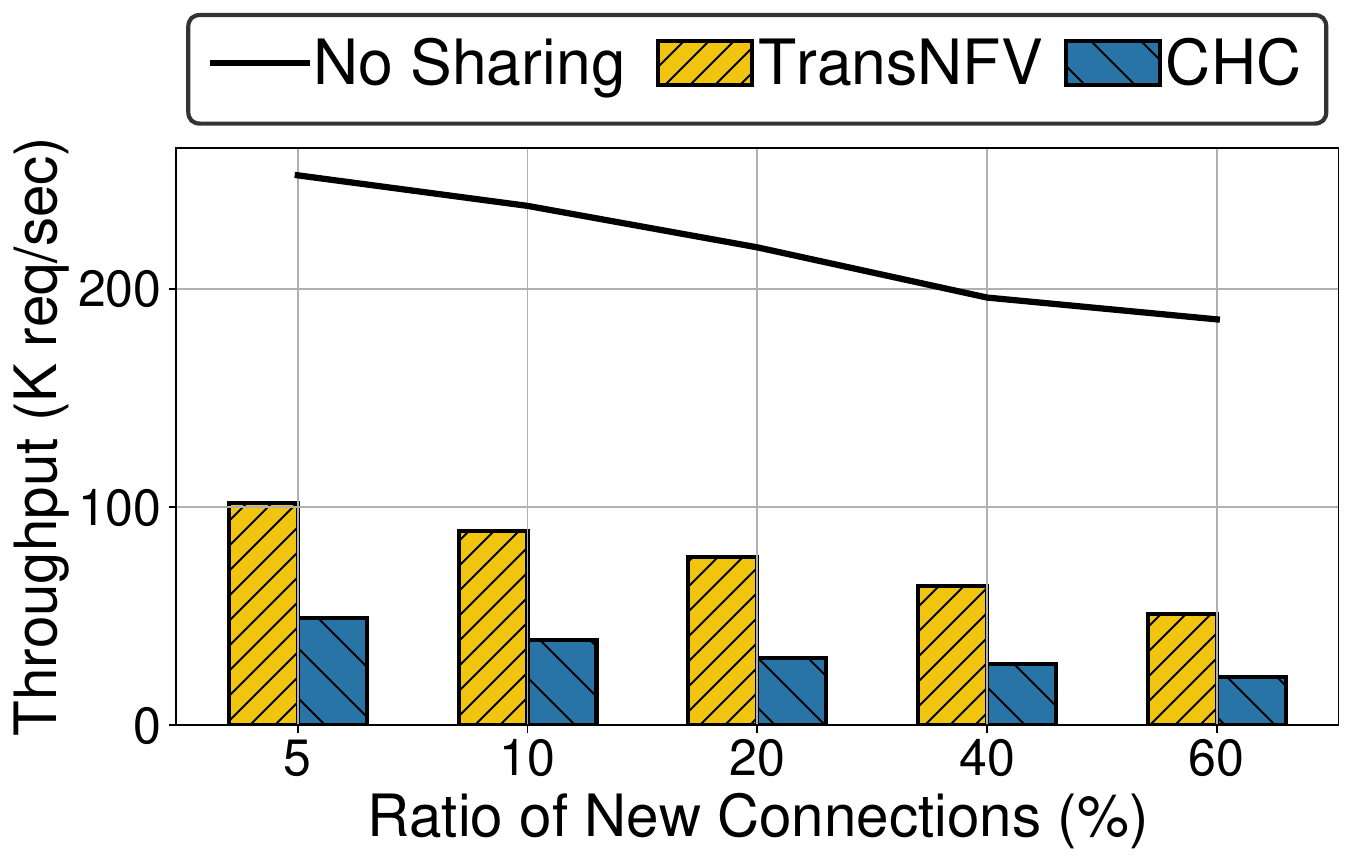}
    \label{fig:Throughput_Ratio}    
}
\subfloat[Latency]{
\includegraphics[width=0.33\textwidth]{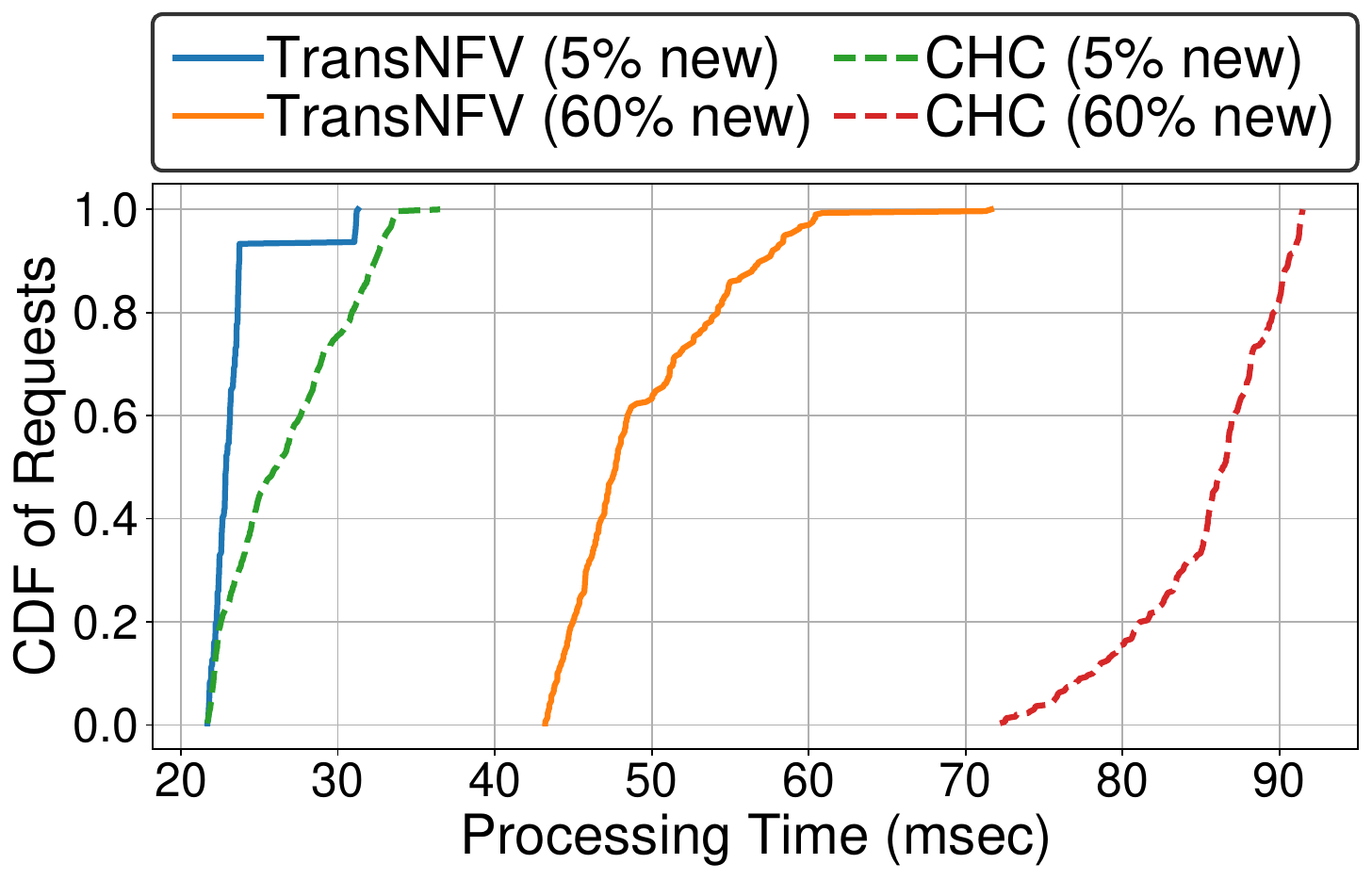}
    \label{fig:Latency_CDF}
}
\subfloat[Execution time breakdown]{
\includegraphics[width=0.33\textwidth]{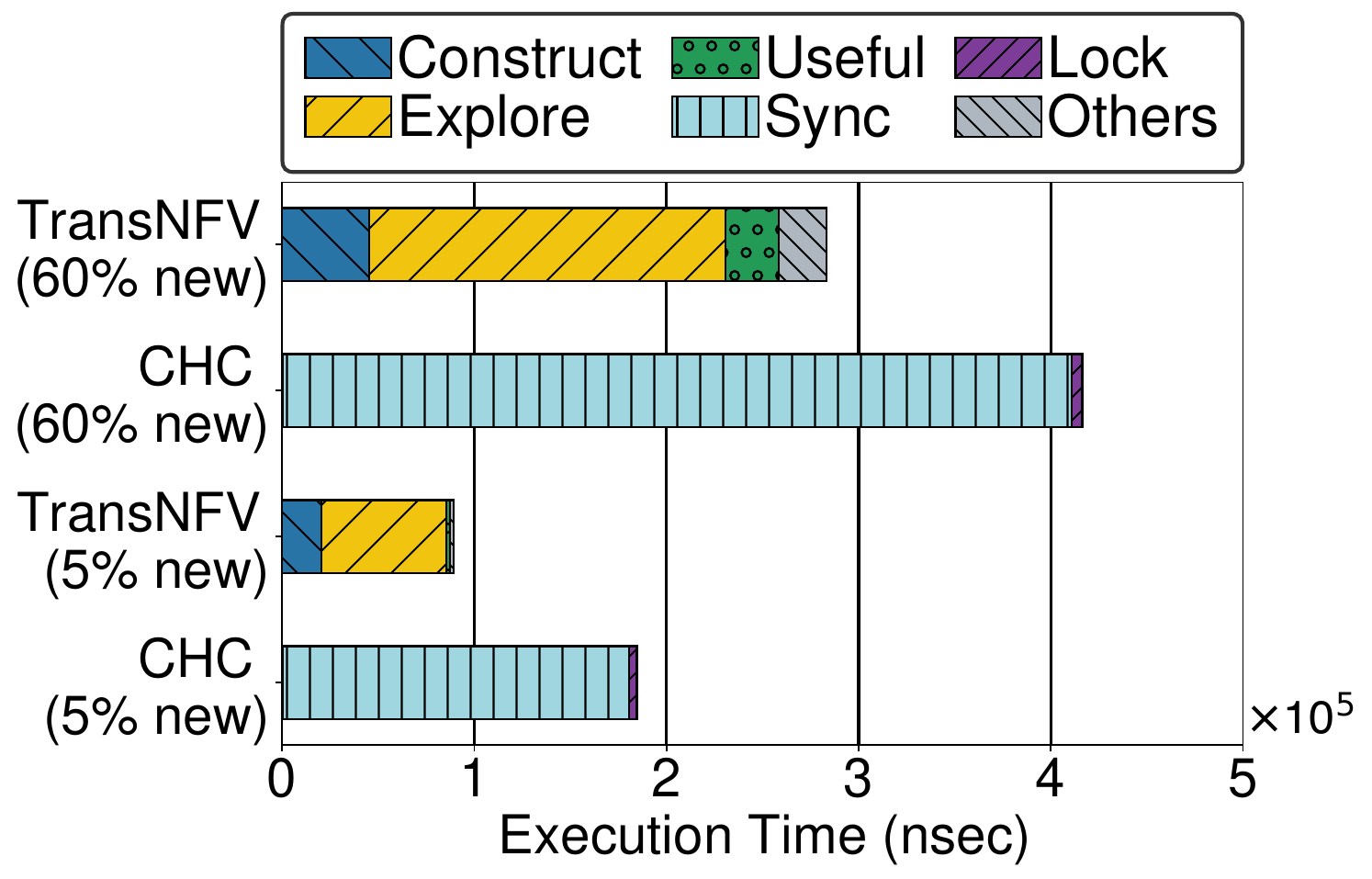}
    \label{fig:Breakdown}
}
\caption{System comparison under various ratios of new connections of LB.}
\end{minipage}
\end{figure*}
\compact
\section{Preliminary Results}
\label{Experiments}
We showcase \system's effectiveness in VNF state management by implementing load balancing (LB) as described by Khalid et al.~\cite{khalid2019correctness}, which is part of our motivating example discussed in Section~\ref{sec:motivation}. We leave the evaluation of the entire NFV chain as future work.
The implementation of LB employs two state tables: a connection counter table, and a connection server mapping table. The LB instance uses these tables to identify new connections, find the server with the fewest active connections, update the relevant counter value, and direct the packet to the correct server. 
We have purposefully designed the input network traffic to include a significant proportion of packets from new connections, all contending to establish a link with the shared server. In contrast to packets from established connections that are processed without requiring state access, each packet from a new connection triggers a series of state access to pinpoint the least busy server and subsequently update the load counter of the selected server.
Simultaneously, other instances are tasked with managing new connections, which are likely to access overlapping ranges of server states. This situation poses a significant challenge in efficiently identifying appropriate servers without compromising the requirements of correctness. This provides a stringent test for \system's capability to handle high concurrency and contention in state access operations effectively and accurately. 

\subsection{Performance Overview}
\label{subsec:overview}
We evaluate the performance of \system and CHC by comparing three primary performance metrics: throughput, latency, and the breakdown of execution time.
We configured \system to run on a dual-socket Intel Xeon Gold 6248R server with 384 GB DRAM. Each socket contains $24$ cores of 3.00GHz and 35.75MB of L3 cache. The OS kernel is \emph{Linux 4.15.0-118-generic}.

\textbf{Throughput.}
We initially conducted experiments to gauge the influence of varying ratios of new connection requests (i.e., 5\% to 60\%) among all incoming packets on application throughput. The results are presented in Figure~\ref{fig:Throughput_Ratio}. We also studied the system performance when locks are completely removed, denoted by \textit{No-Lock}. This represents an ideal situation and offers an upper limit on the system performance. From this analysis, we observed two main points:
1) With an increasing ratio of new connections, the throughput for both \system and CHC declines due to the addition of more state accesses by new connections. However, \system consistently outperforms CHC by achieving significantly higher throughput, approximately twice as much. This affirms the efficiency of incorporating transactional semantics into shared VNF state access management in \system.
2) Despite \system's superior performance over CHC in shared VNF state management, it exhibits a throughput that is 2-3 times lower compared to the \textit{No-Lock}. This suggests substantial room for optimization in the management of concurrent state access during VNF execution.
\textbf{Latency.}
Subsequently, we conducted experiments to juxtapose the processing latency of \system and CHC, the results of which are shown in Figure~\ref{fig:Latency_CDF}. We primarily measured the processing latency of the Load Balancer (LB) handling new connection requests under two extremes: 5\% and 60\% of new connections. These represented the two poles in our prior experiment. As the ratio of new connections increased, so did the processing time of new requests. However, \system managed to consistently achieve significantly lower processing latency (up to 1.8 times less) compared to CHC. Much like our previous results, this observation further corroborates the efficiency of introducing transactional semantics into shared VNF state access management.

\subsection{Processing Overhead}
\label{subsec:overhead}
To delve further into the LB's processing time, we performed a dissection of the time utilized by each aspect of the VNF state access execution under \system and CHC.

We show the time breakdown in the following perspectives. 
1) \emph{Useful Time} refers to the time spent on actual state access.
2) \emph{Sync Time} refers to the time spent on synchronization, including blocking time before lock insertion is permitted or blocking time due to synchronization barriers during mode switching.
3) \emph{Lock Time} refers to the time spent on inserting locks after it is permitted.
4) \emph{Construct Time} refers to the time spent on constructing the auxiliary data structures, e.g., TPG in \system.
5) \emph{Explore Time} refers to the time spent on exploring available operations to process.
6) \emph{Others} refers to all other operations and system overheads, such as index lookup, context switching, and remote memory access.

Figure~\ref{fig:Breakdown} illustrates the time breakdown per state access in CHC and \system. We mainly have two main observations.
First, although \system dedicates a substantial portion of time to exploration (\emph{Explore Time}), it significantly reduces synchronization (\emph{Sync Time}) and lock (\emph{Lock Time}) overhead compared to CHC. This elucidates its superior performance on multicore processors. Second, \system still dedicates a significant amount of time to exploration (\emph{Explore Time}). Moreover, as the ratio of new connections increases, the fraction of exploration time enlarges. This consequently diminishes throughput and increases processing time, as shown in Section~\ref{subsec:overview}. This phenomenon mainly arises from excessive message-passing between threads to identify available operations, and more dependencies lead to more message-passing overhead.

\compact
\section{Conclusion and Future Work}
\label{conclusion}


This paper introduces a novel Network Function Virtualization (NFV) framework, \system, that incorporates transactional semantics into VNF state management. Our primary goal is to optimize the performance of stateful VNFs while maintaining compliance with COE requirements. To this end, we propose a fine-grained dependency resolution for concurrent state access operations, drawing upon transaction processing techniques. Preliminary experimental evaluations demonstrate that these key mechanisms effectively mitigate contention overhead arising from concurrent state accesses. 

Going forward, we see several promising avenues for further exploration. We aim to improve the efficiency of dependency identification by integrating advanced techniques such as machine learning or static analysis. The interplay between Software-Defined Networking (SDN) and \system may also offer intriguing opportunities to adapt network policies based on workload characteristics. 
Furthermore, while \system currently manages transaction failures, we aim to enhance its resilience to system-wide failures by developing robust checkpointing and recovery mechanisms. 
We look forward to investigating these intriguing research opportunities as we continue our efforts to optimize stateful VNF execution in line with the evolving demands of network infrastructures.

\compact

\section*{Acknowledgments}
This work is supported by the National Research Foundation, Singapore and Infocomm Media Development Authority under its Future Communications Research \& Development Programme FCP-SUTD-RG-2022-005.

\bibliographystyle{abbrv} 
\begin{small}
\bibliography{reference}
\end{small}

\end{document}